\documentclass[twocolumn,showpacs,aps,prl,superscriptaddress,preprintnumbers,amsmath,amssymb,floatfix]{revtex4}
\usepackage{bm}%
\usepackage{graphicx}
\usepackage{amsmath}
\usepackage{epsfig}
\usepackage{booktabs} 
\usepackage{multirow} 
\usepackage{dcolumn} 

\RequirePackage{xspace}
\usepackage{relsize}
\def\to                 {\ensuremath{\rightarrow}\xspace}
\def\babar{\mbox{\slshape B\kern-0.1em{\smaller A}\kern-0.1em
    B\kern-0.1em{\smaller A\kern-0.2em R}}}

\def\pisoftp    {\ensuremath{\pi_{\rm s}^{+}}\xspace}
\def\mKKpipi {\ensuremath{m(\Kp\Km\pip\pim)}\xspace}
\def\dm         {\ensuremath{\Delta m}\xspace}
\mathchardef\Upsilon="7107
\def\Y#1S{\ensuremath{\Upsilon{(#1S)}}\xspace}
\def\FourS {\Y4S}
\def\Dbar    {\kern 0.2em\overline{\kern -0.2em D}{}\xspace}

\def\Dstarp  {\ensuremath{D^{*+}}\xspace}
\def\Ds      {\ensuremath{D^+_s}\xspace}
\def\Dz      {\ensuremath{D^0}\xspace}
\def\Dp      {\ensuremath{D^+}\xspace}
\def\Dzb     {\ensuremath{\Dbar^0}\xspace}
\def\Kp    {\ensuremath{K^+}\xspace}
\def\Km    {\ensuremath{K^-}\xspace}
\def\KS    {\ensuremath{K^0_{\scriptscriptstyle S}}\xspace}
\def\pip   {\ensuremath{\pi^+}\xspace}
\def\pim   {\ensuremath{\pi^-}\xspace}
\def\sys {\hbox{syst}}
\def\sta {\hbox{stat}}
\def\pep2{PEP-II}
\def\invfb   {\ensuremath{\mbox{\,fb}^{-1}}\xspace}
\def\epem       {\ensuremath{e^+e^-}\xspace}
\def\ccbar {\ensuremath{c\overline c}\xspace}
\def\CP                {\ensuremath{C\!P}\xspace}
\def\CPT               {\ensuremath{C\!PT}\xspace} 

\newcommand{\gevc}{\ensuremath{{\mathrm{\,Ge\kern -0.1em V\!/}c}}\xspace}
\newcommand{\mevc}{\ensuremath{{\mathrm{\,Me\kern -0.1em V\!/}c}}\xspace}
\newcommand{\gevcc}{\ensuremath{{\mathrm{\,Ge\kern -0.1em V\!/}c^2}}\xspace}
\newcommand{\mevcc}{\ensuremath{{\mathrm{\,Me\kern -0.1em V\!/}c^2}}\xspace}
\newcommand{\kevcc}{\ensuremath{{\mathrm{\,ke\kern -0.1em V\!/}c^2}}\xspace}

\newcommand{\BABARPubYear}    {09}
\newcommand{\BABARPubNumber}  {039}

\newcommand{\SLACPubNumber} {13996}

\def\figurebox#1#2#3{%
    \def\arg{#3}%
    \ifx\arg\empty
    {\hfill\vbox{\hsize#2\hrule\hbox to #2{\vrule\hfill\vbox to #1{\hsize#2\vfill}\vrule}\hrule}\hfill}%
    \else
    {\hfill\epsfbox{#3}\hfill}%
    \fi}

\begin{document}
\preprint{\babar-PUB-\BABARPubYear/\BABARPubNumber} 
\preprint{SLAC-PUB-\SLACPubNumber} 

\title{
{\large \bf
Search for \CP violation using $T$-odd correlations in $\Dz \to \Kp \Km \pip \pim$ decays} 
}
\author{P.~del~Amo~Sanchez}
\author{J.~P.~Lees}
\author{V.~Poireau}
\author{E.~Prencipe}
\author{V.~Tisserand}
\affiliation{Laboratoire d'Annecy-le-Vieux de Physique des Particules (LAPP), Universit\'e de Savoie, CNRS/IN2P3,  F-74941 Annecy-Le-Vieux, France}
\author{J.~Garra~Tico}
\author{E.~Grauges}
\affiliation{Universitat de Barcelona, Facultat de Fisica, Departament ECM, E-08028 Barcelona, Spain }
\author{M.~Martinelli$^{ab}$}
\author{A.~Palano$^{ab}$ }
\author{M.~Pappagallo$^{ab}$ }
\affiliation{INFN Sezione di Bari$^{a}$; Dipartimento di Fisica, Universit\`a di Bari$^{b}$, I-70126 Bari, Italy }
% ...
\collaboration{The \babar\ Collaboration}
\noaffiliation

\date{\today}

\begin{abstract}
We search for CP violation in a sample of $4.7\times10^4$ 
Cabibbo suppressed $\Dz \to \Kp \Km \pip \pim$ decays.
We use 470~${\rm fb}^{-1}$ of data
recorded by the \babar\  detector at the \pep2 asymmetric-energy $e^+e^-$
storage rings running at center-of-mass energies near 10.6 GeV.
\CP violation is searched 
for in the difference between the $T$-odd asymmetries, obtained using 
triple product correlations, measured for \Dz and \Dzb decays. 
The measured \CP violation parameter is ${\cal A}_T = (1.0 \pm 5.1_{\sta}\pm 4.4_{\sys}) \times 10^{-3}$.

\end{abstract}

\maketitle
In the Standard Model (SM) of particle physics, \CP violation arises from a complex phase in the 
Cabibbo-Kobayashi-Maskawa quark mixing matrix~\cite{CKM}.
Physics beyond the SM, often referred to as New Physics (NP), can manifest itself through the
production of new particles, probably at high mass, or through rare processes
not consistent with SM origins. SM predictions for \CP asymmetries in charm meson decays are generally
of $\mathcal{O}(10^{-3})$, at least one order of magnitude lower than current experimental limits~\cite{CP}.
Thus, the observation of \CP violation with current sensitivities would be a NP signal.
Among all hadronic $D$ decays, singly Cabibbo suppressed decays are uniquely sensitive to \CP violation in 
$c \to u \bar q q$ transitions, effect not expected in Cabibbo favored or doubly Cabibbo suppressed decays~\cite{grossman}. 
 
In this paper we report a search for \CP violation in the decay $\Dz \to \Kp \Km \pip \pim $ using a kinematic
triple product correlation of the form $ C_T = {\bf p_1 \cdot ( {p_2 \times p_3 } ) } $,
where each
$ {\bf p_i} $ is a momentum vector of one of the particles in the decay.
The product is odd under time-reversal ($ T $) and,
assuming the $ \CPT $ theorem, $ T $-violation is a signal for \CP-violation.
Strong interaction dynamics can produce a non-zero value of the $ A_{T} $ asymmetry,
\begin{equation}
 A_{T}  \equiv \frac{\Gamma(C_T>0) - \Gamma(C_T<0)}
                    {\Gamma(C_T>0) + \Gamma(C_T<0)},
\end{equation}
\noindent 
where $\Gamma$ is the decay rate for the process, even if the
weak phases are zero. 
Defining as $\overline{A}_T$ the $T$-odd asymmetry measured in the \CP-conjugate decay process, 
\begin{equation}
\overline{A}_{T}  \equiv \frac{\Gamma(-\overline {C}_T>0) - \Gamma(- \overline {C}_T<0)}
                    {\Gamma(-\overline {C}_T>0) + \Gamma(-\overline {C}_T<0)},
\end{equation}
we can construct:
\begin{equation}
 {\cal A}_T = \frac{1}{2}(A_T-\overline{A_T}),
\end{equation}
which is a true $ T $-violating signal~\cite{bensalem}.
At least four particles are required in the final
state so that the three used to define the triple
product are independent~\cite{golowich} of each other. Singly Cabibbo suppressed decays 
having relatively high branching fractions and four different particles in the final state, therefore suitable
for this type of analysis, are $\Dz \to \Kp \Km \pip \pim $ (explored in this paper) and $\Dp \to \Kp \KS \pip \pim$. 
A full angular analysis of these $D$ decays is suggested as a method for searching for \CP violation~\cite{kang}.  
 
Following the suggestion by I.I.~Bigi~\cite{bigi}
to study \CP violation using this technique,
the FOCUS collaboration made the first
measurements using approximately 800 events and reported 
$  {\cal A}_T(\Dz \to \Kp \Km \pip \pim) = 0.010 \pm 0.057 \pm 0.037 $~\cite{focus}.
We perform a similar study using approximately $4.7 \times 10^4$ events.

This analysis is based on
 a 470~$\invfb$ data sample recorded at the
\FourS resonance and 40 MeV below the resonance by the \babar\ detector at the \pep2
asymmetric-energy $e^+e^-$ storage rings.  
The \babar\ detector is
described in detail elsewhere~\cite{babar}. We mention here only the parts of the 
detector which are used in the present analysis.
Charged particles are detected
and their momenta measured with a combination of a 
cylindrical drift chamber (DCH)
and a silicon vertex tracker (SVT), both operating within the
1.5 T magnetic field of a superconducting solenoid. 
The information from
a ring-imaging Cherenkov detector combined with energy-loss measurements in the 
SVT and DCH provide identification of charged kaon and pion candidates. 

The reaction~\cite{conj} 
\begin{equation}
e^+ e^- \to  X \ \Dstarp; \ \Dstarp\to \pip_s \Dz; \  \Dz \to  \Kp \Km \pip \pim,
\end{equation}
\noindent
where $X$ indicates any system composed by charged and neutral particles, has been reconstructed from the sample of 
events having at least five charged tracks. We first reconstruct the $\Dz$ candidate. 
All $\Kp \Km \pip \pim$ combinations assembled from well-measured and 
positively identified kaons and pions are constrained to a common vertex requiring a $\chi^2$ fit probability 
greater than 0.1\%. 
To reconstruct the \Dstarp candidate,
we perform a vertex fit of the $\Dz$ candidates
with all combinations of charged tracks having a laboratory momentum below 0.65 \gevc $(\pisoftp)$ with the 
constraint that the
new vertex is located in the interaction region. We require the fit probability to 
be greater than 0.1\%.
\begin{figure*}[!htb]
\begin{center}
\includegraphics[width=18cm]{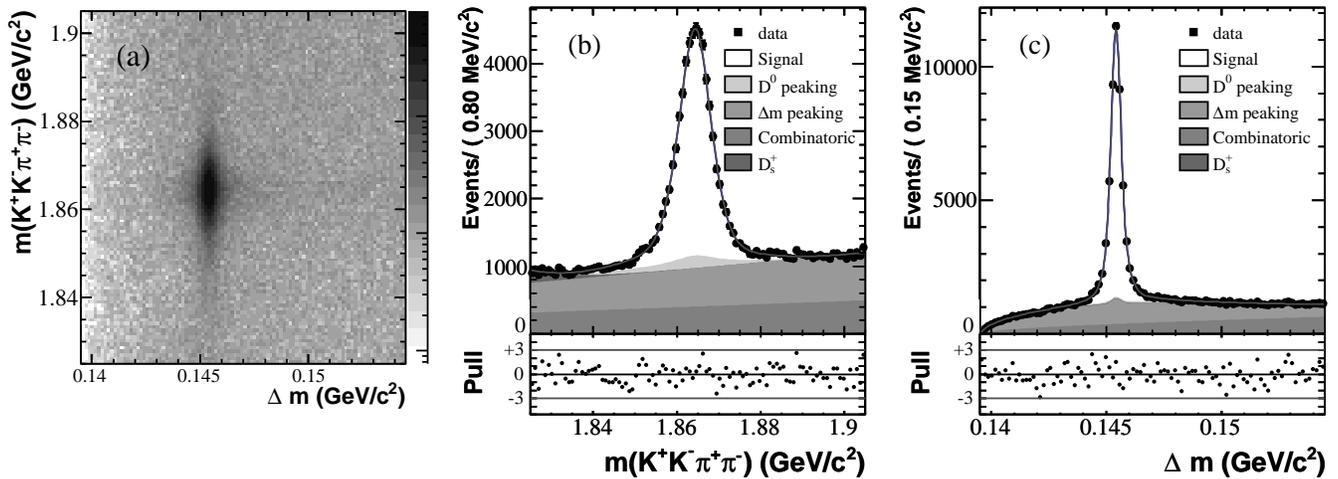}
\caption{(a) $ m(\Kp \Km \pip \pim)$ vs. $\Delta m$ for the total data sample. (b) \mKKpipi and (c) \dm projections with 
curves from the fit results.
Shaded areas indicate the different contributions. The fit residuals, represented by the pulls, are also shown under each distribution.}
\label{fig:fig1}
\end{center}
\end{figure*} 

We require the $\Dz$ to have a center-of-mass momentum greater than 2.5\gevc.
This requirement removes any $\Dz$ coming from $B$ decays.
We observe a contamination of the signal sample from $\Dz \to \Kp \Km \KS$, 
where $\KS \to \pip \pim$. The $\pip \pim$ effective mass shows, in fact, a distinct $\KS$ 
mass peak,
which can be represented by a Gaussian distribution with $\sigma=4.20 \pm 0.26$ $\mevcc$, and which accounts for $5.2\%$ of the selected data sample. 
We veto $\KS$ candidates within a window of 2.5 $\sigma$.
This cut, while reducing to negligible level the background from \Dz~\to~\Kp~\Km~\KS, removes $5.8\%$ of the signal events. 

We look for backgrounds from charm decay modes with mis-identified pions
by assigning alternatively the pion mass to both kaons. Then we study 
the two-body, three-body, four-body and five-body mass distributions (including the $\pisoftp$). We observe a signal 
of $\Ds \to \Kp \Km \pip \pim \pisoftp$ in the five-particle mass distribution, which is taken into account in 
the following fit. 
No other signal is observed in the resulting mass spectra.

We define the mass difference $\Delta m$ as:
\begin{equation}
\Delta m  \equiv m(\Kp \Km \pip \pim \pisoftp) -  m(\Kp \Km \pip \pim).
\end{equation}
\noindent
Figure~\ref{fig:fig1}(a) shows the scatter plot $ m(\Kp \Km \pip \pim)$ vs. $\Delta m$  for all the events.
Figure~\ref{fig:fig1}(b) shows the $ m(\Kp \Km \pip \pim)$ projection, Fig.~\ref{fig:fig1}(c) shows 
the $\Delta m$ projection. 

We perform a fit to the $ m(\Kp \Km \pip \pim)$ and $\Delta m$ distributions,
using a polynomial background and a single Gaussian.
The fit gives $\sigma_{\Dz}=3.94 \pm 0.05\ \mevcc$ for the $\Dz$ mass and 
$\sigma_{\Dstarp}=244 \pm 20  \kevcc$ for the $\Delta m$. We define the signal region within 
$\pm 2  \sigma_{\Dz}$ and $\pm 3.5  \sigma_{\Dstarp}$.
The total yield of tagged $\Dz$ mesons in the signal region is approximately $4.7 \times 10^4$ events. 

The $\Dz$ yields to be used in the calculation of the $T$ asymmetry
are determined using a binned, extended
maximum-likelihood fit to the 2-D (\mKKpipi, \dm) distribution obtained with the two
observables \mKKpipi and \dm in the mass regions defined in the ranges $1.825<\mKKpipi<1.915 \ \gevcc$ and $0.1395<\dm<0.1545 \ \gevcc$ respectively.
Events having more than
one slow pion candidate in this mass region are removed (1.8 \% of the final sample).
The final 2-D distribution contains approximately $1.5 \times 10^5$ events
and is divided into a $100 \times 100$ grid.  
 
The 2-D (\mKKpipi, \dm) distribution is described by five components:
\begin{enumerate}
\item True \Dz signal originating from a \Dstarp decay. This component has characteristic peaks in both observables \mKKpipi and \dm.
\item Random $\pisoftp$ events where a true \Dz is associated to an incorrect $\pisoftp$, called \Dz peaking.
This contribution has the same shape
in \mKKpipi as signal events, but does not peak in \dm.
\item Misreconstructed \Dz decays where one or more of the \Dz decay products are either not
reconstructed or reconstructed with the wrong particle
hypothesis, called \dm peaking. Some of these events show a peak in \dm, but not in \mKKpipi.
\item Combinatorial background where the \Kp, \Km, \pip, \pim candidates are not fragments of the same \Dz decay, called combinatoric.  
This contribution does not exhibit any peaking structure in
\mKKpipi or \dm. 
\item $\Ds \to \Kp \Km \pip \pim \pip$ contamination, called \Ds. 
This background has been studied on Monte Carlo (MC) simulations and shows a characteristic linear narrow shape in the 2-D (\mKKpipi, \dm) distribution,
too small to be directly visible in Fig.~\ref{fig:fig1}(a).
\end{enumerate}

The functional forms of the probability density functions (PDFs) for
the signal and background components are based on studies of
MC samples. These events are generated using the
\textsc{Geant4} program~\cite{geant} and are processed through the same
reconstruction and analysis chain as the real events.
However, all parameters related to these functions are determined from
two-dimensional likelihood fits to data over the full \mKKpipi vs.\ \dm
region. We make use of combinations of Gaussian and Johnson SU~\cite{jsu} lineshapes for peaking distributions, and we use polynomials and
threshold functions for the non-peaking backgrounds. 

The event yields and fractions of
the different components arising from the fit are given in Table~\ref{tab:fit} and shown in Fig.~\ref{fig:fig1}.
The fit residuals shown under each distribution are represented by $Pull=(N_{data} - N_{fit})/\sqrt{N_{data}}$.
\begin{table}[ht]
\caption{Fitted number of events for each category.}
\label{tab:fit}
\begin{center}
\begin{tabular}{lcc}
\hline
Category & Events & \ Fraction (\%)\\
\hline
1. Signal                             & $ 46691 \pm  241$ & $30.8 \pm 0.3$ \\
2. \Dz peaking                 & $   5178 \pm  331$ & $  3.4 \pm 0.2$ \\
3. \dm peaking    & $ 57099 \pm  797$ & $37.7 \pm 0.6$ \\
4. Combinatoric                 & $ 40512 \pm  818$ & $26.7 \pm 0.6$  \\
5. \Ds              & $   2023 \pm  156$ & $ 1.3 \pm 0.1$ \\
\hline
Total                                    & $151503 \pm  1223$&\\
\hline
\end{tabular}
\end{center}
\end{table}

Using momenta of the decay particles calculated in the 
\Dz rest frame, we define the triple product correlations $C_T$ and $\overline {C}_T$ as
\begin{equation}
\begin{split}
 C_{T}  \equiv \vec{p}_{\Kp}\cdot(\vec{p}_{\pip}\times\vec{p}_{\pim}), \\
 \overline {C}_{T}  \equiv \vec{p}_{\Km}\cdot(\vec{p}_{\pim}\times\vec{p}_{\pip}).
\end{split}
\end{equation}

\begin{figure}[!htb]
\begin{center}
\includegraphics[width=9cm]{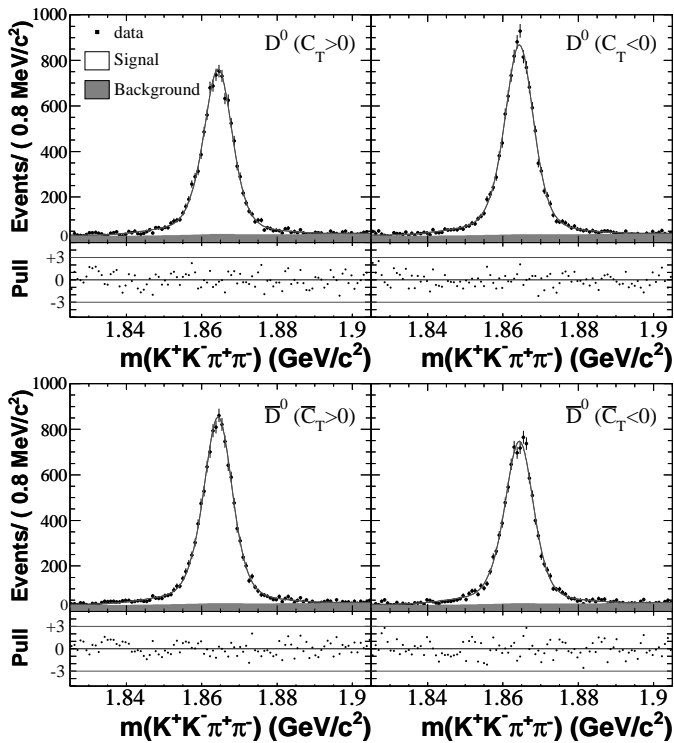}
\caption{Fit projections onto the $m(\Kp \Km \pip \pim)$ for the four different $C_T$ subsamples with cut on $\Delta m$. 
The shaded areas indicate the total backgrounds. The fit residuals, represented by the pulls are also shown under each distribution.}
\label{fig:fig2}
\end{center}
\end{figure}

According to the \Dstarp tag and the $C_T$ variable, we divide the total data sample into four subsamples, defined in Table~\ref{tab:ct}.
These four data samples are fit with fixed PDFs from the total sample. 
The signal event yields are given in Table~\ref{tab:ct}.
Fig.~\ref{fig:fig2} shows the \Kp~\Km~\pip~\pim mass distributions for the four different $C_T$ subsamples 
with fit projections in the $\Delta m$ 
signal region previously defined. 

\begin{table}[!htb]
\caption{Definition of the four subsamples and the event yields from the fit.}
\begin{center}
\begin{tabular}{c|c}
\hline
Subsample & Events \cr
\hline
(a) \Dz, $C_T>0$ &  10974 $\pm$ 117 \cr
(b) \Dz, $C_T<0$ & 12587 $\pm$ 125  \cr
(c) \Dzb, $\overline {C}_T>0$ & 10749 $\pm$ 116  \cr
(d) \Dzb, $\overline {C}_T<0$ &  12380 $\pm$ 124 \cr
\hline
\end{tabular}
\label{tab:ct}
\end{center}
\end{table}

We validate the method using \epem \to \ccbar MC simulations, where \Dz  
decays through the intermediate resonances with the branching  
fractions reported in the PDG~\cite{pdg}. We obtain a T asymmetry 
$\mathcal{A}_T=(2.3\pm 3.3)\times 10^{-3}$, consistent with the generated  
value of $1.0\times 10^{-3}$.

To test the effect of possible asymmetries generated by the detector, we use signal MC 
in which the \Dz decays uniformly over phase space.
In this case possible asymmetries are generated only by the detector efficiency.
These reconstructed events give an asymmetry ${\cal A}_T=−(1.1 \pm 1.1) \times 10^{-3}$,
again consistent with zero.

To avoid potential bias, all event selection criteria are determined
before separating the data into the four subsamples of Table~\ref{tab:ct}.
Systematic uncertainties are obtained directly from the data. In these studies
the true $A_T$ and $\overline{A}_T$ central values are masked by adding unknown random offsets.

After removing the offsets, we measure the following asymmetries:
\begin{equation}
\begin{split}
A_T = (- 68.5 \pm 7.3_{\sta} \pm 5.8_{\sys}) \times 10^{-3}, \\
\overline{A}_T =  (- 70.5 \pm 7.3_{\sta} \pm 3.9_{\sys}) \times 10^{-3}.
\end{split} 
\end{equation}

We observe non-zero values of $A_T$ and $\overline{A}_T$ indicating that final state interaction
effects are significant in this \Dz decay. No effect is found, on the other hand, in the analysis of MC samples.
Final state interactions effects are common in hadronic $D$ decays because of the complex interference patterns
between intermediate resonances formed between hadrons in the final states~\cite{oller}.

The result for the \CP violation parameter, ${\cal A}_T$, is
\begin{equation}
{\cal A}_T = (1.0 \pm 5.1_{\sta}\pm 4.4_{\sys}) \times 10^{-3}. 
\end{equation}

The sources of systematic uncertainties considered in this analysis are listed in Table~\ref{tab:sys}.
The estimates of their values are derived as follows:
\begin{enumerate}
\item The PDFs used to describe the signal are modified, replacing the Johnson SU function by a Crystal Ball
function~\cite{cb}, obtaining fits of similar quality.
\item As in 1.,  for the peaking background.
\item We increase the number of bins of the 2-D (\mKKpipi, \dm) distribution to a ($120 \times 120)$ grid and decrease to a grid of $(80 \times 80)$. 
\item The particle identification algorithms used to identify kaons and pions are modified to more 
stringent conditions in different combinations. We notice that the difference between different selection 
efficiencies is significantly larger than the uncertainties on efficiency of  
the default selection. On the other hand, the use of the discrepancy between data and MC obtained using high statistics control samples, gives a much lower contribution.
\item The $p^*(\Dz)$ cut is increased to 2.6\gevc and 2.7\gevc.
\item We study possible intrinsic asymmetries due to the interference
between the electromagnetic $e^+e^-\to \gamma^* \to \ c\bar{c}$ and weak
neutral current $e^+e^-\to Z^0\to \ c\bar{c}$ amplitudes. 
This interference produces a \Dz/\Dzb production asymmetry
that varies linearly with the quark production angle with respect to the
$e^-$ direction. Since \babar\ is an asymmetric detector, the final yields
of \Dz and \Dzb are not equal. We constrain the possible 
systematics by measuring $\mathcal{A}_T$ in three regions of 
the center-of-mass \Dz production angle $\theta^*$: forward ($0.3 < \cos(\theta^*)_{\Dz}$), 
central ($-0.3 < \cos(\theta^*)_{\Dz} \leq 0.3$), and backward ($ \cos(\theta^*)_{\Dz} < -0.3$). 
We observe that the $\mathcal{A}_T$ angular variation is, within the large statistical errors, 
consistent with zero as expected from the MC
\item Fit bias: we use MC simulations to compute the difference between the generated and reconstructed ${\cal A}_T$.
\item Mistag: there are a few ambiguous cases with more than one $D^*$ in the event. We use MC simulations where
these events are included or excluded from the analysis. This effect has a negligible contribution to the systematic
uncertainty. 
\item Detector asymmetry: we use the value obtained from the MC simulation where \Dz decays uniformly 
over the phase space.  
\end{enumerate}
In the evaluation of the systematic uncertainties, we keep, for a given category, 
the largest deviation from the reference value
and assume symmetric uncertainties. Thus, most systematic uncertainties have a statistical component,
and are conservatively estimated.
\begin{table}[ht]
\caption{Systematic uncertainty evaluation on ${\cal A}_T$, ${A}_T$, and $\overline{A}_T$ in units of $10^{-3}$.}
\label{tab:sys}
\begin{center}
\begin{tabular}{lccc}
\hline \noalign{\vskip2pt}
Effect & ${\cal A}_T$ & ${A}_T$ & $\overline{A}_T$\\
\hline
1. Alternative signal PDF                                & $0.2$ & $0.3$ & $0.2$\\
2. Alternative misreconstructed \Dz PDF& $0.5$ & $0.1$ & $0.9$\\
3. Bin size                                                             & $0.2$ & $0.4$ & $0.3$\\
4. Particle identification                                    & $3.5$ & $4.2$ & $2.9$\\ 
5. $p^*(\Dz)$ cut                                                 & $1.7$ & $1.6$ & $2.4$\\
6. $\cos \theta^*$ dependence                 & $0.9$ & $0.0$ & $0.2$\\
7. Fit bias                                                                & $1.4$ & $3.0$ & $0.3$\\
8. Mistag                                                         & $0.0$ & $0.0$ & $0.0$\\
9. Detector asymmetry                                 & $1.1$ & $2.1$ & $0.0$\\
\hline
Total                                                                & $4.4$ & $5.8$ & $3.9$\\
\hline
\end{tabular}
\end{center}
\end{table}
In conclusion, we search for \CP violation using $T$-odd correlations in a high statistics sample of Cabibbo suppressed 
$\Dz \to \Kp \Km \pip \pim$ decays. We obtain a $T$-violating asymmetry consistent with zero with a 
sensitivity of $\approx$ 0.5 \%. 

The study of triple product correlations in $B$ decays shows evidence for final state interaction but 
also give asymmetries consistent with zero, in agreement with SM expectations~\cite{bphys}.
These results constrain the possible effects of New Physics in this observable~\cite{grossman}. The results 
from this analysis fix a reference point, since the study of $T$-odd correlations play an important role in the Physics 
program of present and future charm and $B$-factories.

We are grateful for the excellent luminosity and machine conditions
provided by our \pep2\ colleagues, 
and for the substantial dedicated effort from
the computing organizations that support \babar.
The collaborating institutions wish to thank 
SLAC for its support and kind hospitality. 
This work is supported by
DOE
and NSF (USA),
NSERC (Canada),
CEA and
CNRS-IN2P3
(France),
BMBF and DFG
(Germany),
INFN (Italy),
FOM (The Netherlands),
NFR (Norway),
MES (Russia),
MEC (Spain), and
STFC (United Kingdom). 
Individuals have received support from the
Marie Curie EIF (European Union) and
the A.~P.~Sloan Foundation.


\begin{thebibliography}{99}
\bibitem{conj}
Charge-conjugate reactions are implied throughout.
\end{thebibliography}
\end{document}